\documentstyle[amssymb,aps,prl]{revtex}

\begin{document}
\title{Direct measurement of polaron binding energy in AMnO$_{3}$ as a function of
the A site ionic size by photoinduced IR absorption}
\author{T. Mertelj$^{1,2}$, M. Hrovat$^{1}$, D. Ku\v{s}\v{c}er$^{1}$ and D.
Mihailovic$^{1}$}
\address{$^{1}$Jozef Stefan Institute, P.O.Box 3000, 1001 Ljubljana, Slovenia}
\address{$^{2}$University of Ljubljana, Faculty of Mathematics and Physics,\\
Jadranska~19, 1000 Ljubljana, Slovenia}
\maketitle

\begin{abstract}
Photoinduced IR absorption was measured in undoped (LaMn)$_{1-\delta }$O$%
_{3} $ and (NdMn)$_{1-\delta }$O$_{3}$. We observe broadening and a $\sim $%
44\% increase of the midinfrared anti-Jahn-Teller polaron peak energy when La%
$^{3+}$ is replaced with smaller Nd$^{3+}$. The absence of any concurent
large frequency shifts of the observed PI phonon bleaching peaks and the
Brillouin-zone-center internal perovskite phonon modes measured by Raman and
infrared spectroscopy indicate that the polaron peak energy shift is mainly
a consequence of an increase of the electron phonon coupling constant with
decreasing ionic radius $\left\langle r_{A}\right\rangle $ on the perovskite
A site. This indicates that the dynamical lattice effects strongly
contribute to the electronic band narrowing with decreasing $\left\langle
r_{A}\right\rangle $ in doped giant magnetoresistance manganites.
\end{abstract}

The physical properties of manganites with the chemical formula (Re$_{1-x}$Ae%
$_{x}$)MnO$_{3}$ (Re and Ae are trivalent rare-earth and divalent
alkaline-earth ions respectively) in which giant magnetoresistance (GMR) is
observed\cite{SearleWang69,KustersSingelton89,HelmoltWecker93} show
remarkable changes when the average ionic radius $\left\langle
r_{A}\right\rangle $ on the perovskite A site is varied.\cite
{ImadaFujimori98,HwangCheong95} In the region of doping $x$, where GMR is
observed, this is reflected in a decrease of the Curie temperature $T_{C}$
and increase of the size of magnetoresistance with decreasing $\left\langle
r_{A}\right\rangle $.\cite{HwangCheong95} The decrease of $T_{C}$ has been
attributed to a decrease of the hopping matrix element between neighbouring
Mn sites $t$ as a result of changes of Mn-O-Mn bond angles with $%
\left\langle r_{A}\right\rangle $.\cite{HwangCheong95}

Traditionally GMR has been explained in the double exchange picture\cite
{Zener51} framework, where the hopping matrix element is one of the key
parameters influencing directly the Curie temperature. However it has been
shown experimentally\cite{ZhaoConder96,ZhaoKeller98,LoucaEgami97} and
theoretically\cite{MillisShraiman96} that also dynamic lattice effects
including Jahn-Teller (JT) polaron formation are crucial ingredients for the
explanation of GMR in manganites\cite{MillisShraiman96}. In this picture $%
T_{C}$ also strongly depends on the electron-phonon (EP) coupling in
addition to the hopping matrix element $t$ and any change in the EP coupling
as function of $\left\langle r_{A}\right\rangle $ contributes to changes of $%
T_{C}$ and other physical properties. Experimentally an increase of the EP
coupling with decreasing $\left\langle r_{A}\right\rangle $ is suggested by
the shift of the $1$-eV polaronic peak in optical conductivity of manganites
to higher energy with decreasing $\left\langle r_{A}\right\rangle $\cite
{MachidaMoritomo98,QuijadaCerne98}. Unfortunately, the peak position of the
1-eV peak does not depend on the polaron binding energy alone\cite
{QuijadaCerne98} and the magnitude of the shift can not be directly linked
to change of the EP coupling constant $g$.

Recently we observed a polaronic photoinduced (PI) absorption peak in
antiferromagnetic (LaMn)$_{1-\delta }$O$_{3}$ (LMO).\cite{MerteljKuscer00}
In this case the peak position is directly linked to the anti-Jahn-Teller
polaron\cite{AllenPerebeinos99} binding energy and enables us to {\em %
measure directly} the change of the electron-phonon coupling with $%
\left\langle r_{A}\right\rangle $ in undoped GMR manganites. Here we present
photoinduced (PI) absorption measurements in (NdMn)$_{1-\delta }$O$_{3}$
(NMO) with $\delta \approx 0$. We observe a $\sim $44\% increase of the
small polaron energy when La$^{3+}$is replaced by smaller Nd$^{3+}$. The
absence of any concurrent large frequency shifts of the observed PI phonon
bleaching peaks and the Brillouin-zone-center internal perovskite phonon
modes measured by Raman and infrared (IR) spectroscopy indicate that the
polaron energy increase with decreasing $\left\langle r_{A}\right\rangle $
is mainly a consequence of an increase of the electron-phonon coupling
constant.

The method of preparation and characterization of ceramic sample with
nominal composition (LaMn)$_{1-\delta }$O$_{3}$ has been published elsewhere%
\cite{MerteljKuscer00,HolcKuscer97}. The sample with nominal composition
(NdMn)$_{1-\delta }$O$_{3}$ was prepared in a similar manner with equal
final treatment at 900${{}{}^{\circ }}$C for 300 min in Ar flow\cite
{HuangSantoro97} to decrease cation deficiency. The X-ray difraction
patterns of both samples taken before Ar treatment in 2$\Theta $ range 20${%
{}^{\circ }}$-70${{}^{\circ }}$ showed that both samples are single phase.
The samples showed no sign of a ferromagnetic transition in AC
susceptibility measurements and we concluded that $\delta $ is sufficiently
small that both are antiferromagnetic (AFM) and insulating below their
respective Neel temperatures\cite
{ImadaFujimori98,HuangSantoro97,UrushibaraMoritomo95}.

PI spectra were measured at 25K in samples dispersed in KBr pellets. CW Ar$%
^{+}$-ion-laser light with 514.5 nm wavelength ($h\nu =2.41$ eV) and optical
fluence $\sim $500 mW/cm$^{2}$ was used for photoexcitation. Details of
PI-transmitance spectra measurements were published elswhere.\cite
{MerteljKuscer00} Thermal difference spectra\cite{MerteljKuscer00} (TD) were
also measured at the same temperature eliminate possible laser heating
effects. Raman spectra were measured at room temperature in a standard
backscatering configuration from ceramic powders using a CW Kr$^{+}$%
-ion-laser light at 647.1 nm. The scattered light was analysed with a SPEX
triple spectrometer and detected with a Princeton Instruments CCD array. The
incident laser flux was kept below $\sim $400 W/cm$^{2}$ to avoid laser
annealing.\cite{IlievAbrashev98}

The low temperature ($T=25$K) PI transmittance $(\frac{\Delta {\cal T}_{PI}}{%
{\cal T}})$ spectra of both samples are shown in Fig. 1. In both samples a
strong broad PI midinfrared (MIR) absorption (negative PI transmittance)
centered at $\thicksim $5000 cm$^{-1}$ ($\thicksim 0.62$ eV) in LMO and at $%
\thicksim 7500$ cm$^{-1}$ ($\thicksim 0.93$ eV) in NMO is observed.

In the frequency range of the phonon bands (insert of Fig.1) we observe PI
phonon bleaching in the range of the 585-cm$^{-1}$ (576 cm$^{-1}$ in NMO) IR
phonon band and a slight PI absorption below $\thicksim $580 cm$^{-1}$. The
PI phonon bleaching in NMO is similar to LMO\cite{MerteljKuscer00}, but
shifted to higher frequency by $\thicksim $20 cm$^{-1}$ and it consists of
two peaks at 620 and 690 cm$^{-1}$ with a dip in-between at 660 cm$^{-1}$.
Similarly to LMO this two PI transmission peaks are reproducible among
different runs, while the structure of the PI absorption below $\thicksim $%
580 cm$^{-1}$ is not, and presumably arises due to increasing instrumental
noise at the lower end of the spectral range. Despite the noise a slight PI
absorption below $\thicksim $580 cm$^{-1}$ can be deduced from the PI
spectra.

The Raman spectra shown in Fig. 2b are consistent with published data.\cite
{IlievAbrashev98} In the 100-900-cm$^{-1}$ frequency range 5 phonon peaks
are observed in LMO and 6 phonon peaks in NMO. The frequencies and
assignments of the phonon peaks are shown in Table I. The only mode that
shifts substantially is the $A_{g}$ mode that corresponds to the out of
phase rotation of the MnO$_{3}$ octahedra.\cite{IlievAbrashev98} The mode
frequency increases by 17\% to 329 cm$^{-1}$ in NMO.\cite{ymno} The two high
frequency modes that are expected to be related to the collective JT
distortion\cite{LiarokapisLeventouri99} shift by less than 5 cm$^{-1}$,
which is less than 1 \%. Similarly, the frequencies of IR modes shown in
Table II do not shift more than 6 \% when La$^{3+}$ is replaced with Nd$%
^{3+} $.

A fit of absorption due to a small polaron hopping given by Emin \cite
{Emin93} to the data is shown in Fig. 1 for both samples assuming that $-%
\frac{\Delta {\cal T}_{PI}}{{\cal T}}$ is proportional to the absorption
coefficient\cite{KimHeeger87}: 
\begin{equation}
-\frac{\Delta {\cal T}_{PI}}{{\cal T}}\varpropto \alpha \varpropto \frac{1}{%
\hbar \omega }\exp (-\frac{(2E_{pol}-\hbar \omega )^{2}}{4E_{pol}\hbar
\omega _{ph}})  \label{eqemin}
\end{equation}
where $\alpha $ is the absorption coefficient, $E_{pol}$ is the polaron
binding energy, $\omega $ the incoming photon frequency and $\omega _{ph}$
the polaron phonon frequency. The theoretical prediction fits well to the
data with the small polaron binding energies $E_{pol}=0.34$ eV and $%
E_{pol}=0.49\,$eV in LMO and NMO samples respectively. The polaron binding
energies $E_{pol}$ and polaron phonon frequencies $\omega _{ph}$ obtained
from the fit are summarized in Table III.

The polaron phonon frequencies $\omega _{ph}$ obtained from the fit are 310
and 330 cm$^{-1}$ in LMO and NMO respectively and are small compared to the
frequencies of the JT related Raman modes at $\thicksim $ 490 cm$^{-1}.$
This discrepancy is not surprising, since the width of the peak in (\ref
{eqemin}), from which $\omega _{ph}$ is determined, includes a prefactor
which strongly depends on the details of the phonon cloud in the small
polaron and is 4 only in a 1D Holstein model, which is a molecular model
with a single dispersionless phonon. Taking into account the prefactor as
well as dispersion and multiplicity of the phonon branches, $\omega _{ph}$
can be viewed as an {\em effective polaron phonon frequency} of the
different wavevector phonons in the phonon cloud.

The shape of the observed PI absorption peak indicates that the EP coupling
is strong and $t>\omega _{ph}$, since otherwise one does not expect to
observe a symmetric and structureless peak in optical conductivity.\cite
{AlexandrovKabanov94} In this case the polaron binding energy is
proportional to\cite{KabanovMashtakov93} 
\begin{equation}
E_{pol}\varpropto 2g^{2}\omega _{ph}+\frac{zt^{2}}{2g^{2}\omega _{ph}}\text{.%
}  \label{epol}
\end{equation}
Here $g$ is the dimensionless EP coupling constant, $t$ the bare hopping
matrix element and $z$ the number of nearest neighbours. In this formula
again $g$ and $\omega _{ph}$ should be viewed as effective quantities
corresponding to the combination of different wavevector phonons in the
polaron phonon cloud. From formula (\ref{epol}) it is evident that $E_{pol}$
is very sensitive to changes of the EP coupling constant $g$ and $\omega
_{ph}$, but depends on the bare hopping matrix element $t$ only in the
second order, since in the strong coupling limit $2g^{2}\omega _{ph}\gg t$.

In our experiment one can see from Raman and IR spectra that, apart from the 
$A_{g}$-mode frequency which corresponds to the out of phase rotation of the
MnO$_{3}$ octahedra, all the observed Brillouin-zone-center-phonon
frequencies shift by no more than a few percent when La$^{3+}$ is replaced
by Nd$^{3+}$. This would at a first sight suggest a link between the $A_{g}$
octahedral rotation mode, especially because the effective polaron phonon
frequencies $\omega _{ph}$ obtained from the fit are very close to the
observed mode frequencies. Hardening of the phonon mode by 17\% and the $%
\sim $44\% increase of observed small polaron energy would in this case
according to (\ref{epol}) imply a $\sim $14\% increase of the EP coupling
constant $g.$

However, as stated above, the effective polaron phonon frequencies obtained
from the fit are extremly inaccurate due to a crudness of the Holstein model
and, since the small-polaron phonon cloud due to its localised nature
includes mainly large wavevector phonons, the effective $\omega _{ph}$ only
weakly depends on the frequency of the zone center phonons. In addition, the
observed PI phonon bleaching peaks, which are expected to be directly
related to the nonzero wavevector phonons forming the polaron phonon cloud,
harden by a mere $\sim $3\%. We therefore suggest that it is {\em very
unlikely }that the frequency shift of the $A_{g}$ mode, which corresponds to
the out of phase rotation of the MnO$_{3}$ octahedra, is directly related to
the observed small polaron binding energy shift.

Instead we attribute the $\sim ${\em 44\% increase of the polaron binding
energy to a }$\sim ${\em 20\% increase of the EP coupling constant }$g$ when
La$^{3+}$is replaced by Nd$^{3+}$and not to a change of $\omega _{ph}$
and/or the bare hopping matrix element $t$. This is supported by a small
shift of the observed PI phonon bleaching peaks and, nevertheless, the
negligible shift of the relevant Brillouine zone center phonon modes,
especialy the Raman-active high frequency ones, which have been shown to be
related to the collective JT distortion\cite{LiarokapisLeventouri99}.

The influence of a decrease of the bare hopping matrix element $t$ with
decreasing $\left\langle r_{A}\right\rangle $ on the polaron binding energy
can be neglected due to its second order nature in (\ref{epol}), since a
decrease of $t$ with decreasing $\left\langle r_{A}\right\rangle $ in (\ref
{epol}) would lead to a decrease of $E_{pol}$, which is the {\em opposite}
to what is experimentally observed.

In conclusion, we observe a $\sim $44\% increase of the anti-Jahn-Teller
polaron binding energy when La$^{3+}$is replaced by smaller Nd$^{3+}$in
undoped GMR manganites. Absence of any concurent large frequency shifts of
the observed PI phonon bleaching peaks and the Brillouin-zone-center
perovskite internal phonon modes indicate that the increase of the polaron
binding energy is a consequence of increasing electron-phonon coupling
strength with decreasing ionic radius on the perovskite A site. This result
can be safely extrapolated to doped manganites as indicated by the shift of
the 1-eV polaronic peak with decreasing $\left\langle r_{A}\right\rangle $
in optical conductivity of GMR manganites\cite
{MachidaMoritomo98,QuijadaCerne98} and increasing isotope effect of $T_{C}$
with decreasing $\left\langle r_{A}\right\rangle $\cite{ZhaoKeller98}.

The decrease of the effective bandwidth resulting in decrease of the Curie
temperature $T_{C}$ and increase of the size of magnetoresistance with
decreasing $\left\langle r_{A}\right\rangle $ is suggested to be, not just
due to the direct influence of Mn-O-Mn bond angles on the bare hopping
matrix element $t$,\cite{HwangCheong95} but also a consequence of the
increasing polaronic band narrowing due to increasing electron phonon
coupling.

{\bf Acknowledgments}

We would like to thank V.V. Kabanov for fruitful discussions.

\begin{table}
\caption{The Raman phonon frequencies.}
\begin{tabular}{llll}
Mode assignment\cite{IlievAbrashev98}& (LaMn)$_{1-\delta }$O$_{3}$  (cm$^{-1}$)& (NdMn)$_{1-\delta }$O$_{3}$  (cm$^{-1}$) & Shift \% \\ \hline
$B_{2g}$ octahedra in-phase stretching & 609& 604& -0.8 \\
$A_g$ octahedra out-of-phase bending& 486& 489& -0.6\\
$B_{2g}$ O1 along $z$ &310& -\\
$A_g$ octahedra out-of-phase rotation&282 &329 & 17\\
laser annealing induced&233 & \\
probably laser annealing induced&-&248\\
probably laser annealing induced&-&234\\
$A_g$ Nd along $x$&-&143\\
\end{tabular}
\end{table}
%

\begin{table}
\caption{The IR phonon band frequencies.}
\begin{tabular}{llll}
(LaMn)$_{1-\delta }$O$_{3}$  (cm$^{-1}$)& (NdMn)$_{1-\delta }$O$_{3}$  (cm$^{-1}$)& Shift \% & Comment\\ \hline
636&640&0.6&shoulder\\
584&577&-1.2\\
510&521&2.2\\
460&482&4.8\\
432&458&6.0\\
418&434&3.8\\
376&392&4.2\\
\end{tabular}
\end{table}
%

\begin{table}
\caption{The small polaron binding energy $E_{pol}$ and the effective polaron phonon frequency $ \omega _{ph}$ as obtained from the fit of absorption due to a small polaron
given by Emin\protect\cite{Emin93}. The perovskite A site ionic radii for 9-fold coordination\protect\cite{Shannon76} ($ r_{A}$) are also given for comparison.}
\begin{tabular}{llll}
sample &$ r_{A} (\AA)  $ &   $E_{pol}$ (eV) & $\omega _{ph}$ (cm$^{-1}$) \\ \hline
(LaMn)$_{1-\delta }$O$_{3}$ & 1.216&  $0.34$ & $ 310$ \\ 
(NdMn)$_{1-\delta }$O$_{3}$  & 1.163& $0.49$ & $ 330$ \end{tabular}
\end{table}
%

\section{Figure Captions}

Figure 1. Photoinduced absorption spectra of (LaMn)$_{1-\delta }$O$_{3}$
(thick solid line) and (NdMn)$_{1-\delta }$O$_{3}$ (dashed-line). The thin
lines represent the fit of equation (\ref{eqemin}) to the data. Inset shows
photoinduced absorption spectra in the region of the phonon bands. The
structure of the PI absorption below $\thicksim $580 cm$^{-1}$ is not
reproducible among different runs, and presumably arises due to increasing
instrumental noise at the lower end of the spectral range.

Figure 2. Infrared (a) and Raman (b) phonon spectra of (LaMn)$_{1-\delta }$O$%
_{3}$ (solid line) and (NdMn)$_{1-\delta }$O$_{3}$ (dashed-line). The Raman
spectrum of (NdMn)$_{1-\delta }$O$_{3}$ is offset vertically for clarity.

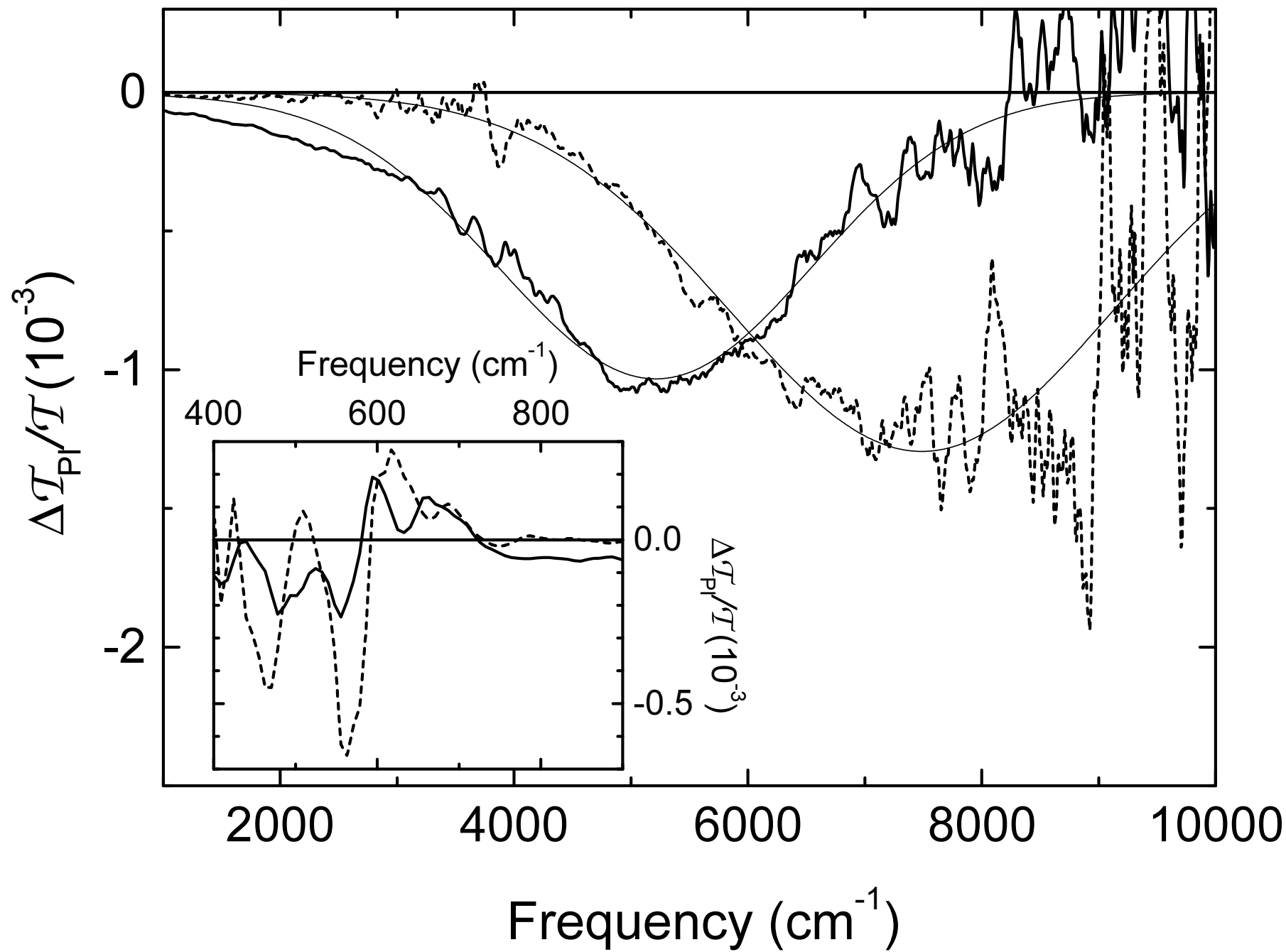
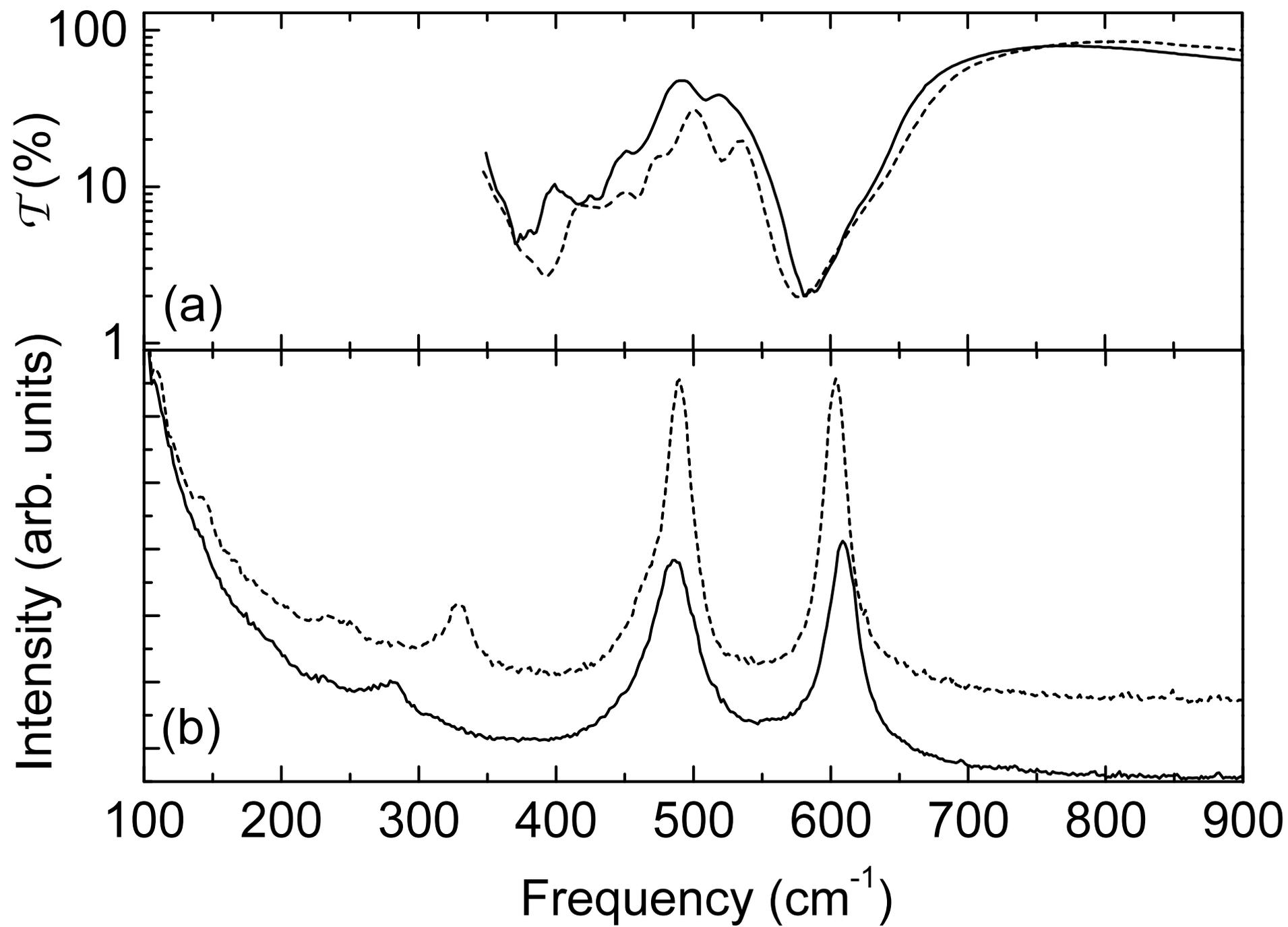
\end{document}